\documentclass[sigconf,nonacm,11pt]{acmart}
\usepackage{enumitem}
\usepackage{tikz}
\usepackage{makecell}
\usepackage{tabularx}
\usepackage{xurl}
\usepackage{hyperref}
\hypersetup{breaklinks=true}
\usetikzlibrary{arrows.meta, positioning, shapes.multipart}
\makeatletter
\def\@ACM@checkaffil{
    \if@ACM@instpresent\else
    \ClassWarningNoLine{\@classname}{No institution present for an affiliation}%
    \fi
    \if@ACM@citypresent\else
    \ClassWarningNoLine{\@classname}{No city present for an affiliation}%
    \fi
    \if@ACM@countrypresent\else
        \ClassWarningNoLine{\@classname}{No country present for an affiliation}%
    \fi
}
\makeatother

\AtBeginDocument{%
  \providecommand\BibTeX{{%
    \normalfont B\kern-0.5em{\scshape i\kern-0.25em b}\kern-0.8em\TeX}}}

\graphicspath{{fig/}{./}}

\acmConference{}
\acmDOI{}
\acmISBN{}
\acmBooktitle{}

\begin{document}

\title{FinMetaMind: A Tech Blueprint on NLQ Systems for Financial Knowledge Search}


\author{Lalit Pant}
\email{lpant3@gatech.edu}
\affiliation{%
Independent Author
}

\author{Shivang Nagar}
\email{shivangnnagar@gmail.com}
\affiliation{%
Independent Author
}




\begin{abstract}
\textbf{Natural Language Query (NLQ) allows users to search and interact with information systems using plain, human language instead of structured query syntax. This paper presents a technical blueprint on the design of a modern NLQ system tailored to financial knowledge search. The introduction of NLQ  not only enhances the precision and recall of the knowledge search compared to traditional methods, but also facilitates deeper insights by efficiently linking disparate financial objects, events, and relationships. Using core constructs from natural language processing, search engineering, and vector data models, the proposed system aims to address key challenges in discovering, relevance ranking, data freshness, and entity recognition intrinsic to financial data retrieval. In this work, we detail the unique requirements of NLQ for financial datasets and documents, outline the architectural components for offline indexing and online retrieval, and discuss the real-world use cases of enhanced knowledge search in financial services. We delve into the theoretical underpinnings and experimental evidence supporting our proposed architecture, ultimately providing a comprehensive analysis on the subject matter. We also provide a detailed elaboration of our experimental methodology, the data used, the results and future optimizations in this study.} 
\end{abstract}



\maketitle
\section{Introduction}
The exponential growth of financial data in today's digital era has created unprecedented challenges for organizations seeking to extract valuable information from vast, complex, and heterogeneous sources. Timely, accurate and contextual access to financial knowledge is essential for informed decision making in domains such as investment analysis, risk management, regulatory compliance, and market monitoring. Traditional keyword-based search systems and unstructured data retrieval mechanisms often fail to capture the nuanced relationships and metadata-rich context that shape financial information.

With the advancement of artificial intelligence, several studies have been conducted to improve the knowledge search capabilities on large enterprise systems. The work published by Bruch \cite{retrieval} on vector retrieval talks about how vectors can improve the search capabilities in large-scale data systems. Further, Montgomery et al. \cite{nlq} describe the design and development of a natural language query interface that proposes a middle layer to store necessary metadata. Finally, Keraghel et al. \cite{ner} talk about advancement in Transformer-based methods and Large Language Models (LLMs) for Named Entity Recognition within text documents. In this paper, we bring together the explorations from these publications and combine them with modern system engineering tools to propose and validate an architecture for NLQ systems applied to the domain of financial technology.

\textbf{TLDR:} This document presents a concise blueprint for  enabling Natural Language Query Systems on financial knowledge repositories. The paper explores the intricate relationship between advanced data science techniques and systems engineering principles within the context of building scalable search systems.

\section{System Overview}

A Natural Language Query System for Financial Knowledge Search enables users to explore complex financial datasets intuitively using human language rather than technical query syntax. Its architecture typically consists of two main components as described in Figure 1: an Offline Indexing Service and an Online Retrieval Service.

The Offline Indexing Service pre-processes and structures the financial documents and metadata, such as transaction records, account dimensions, and regulatory documentation from a system of record into a unified schema. It employs a vector store that encodes this data into dense numerical vectors using transformer-based embeddings like BERT\cite{bert} or OpenAI’s text embedding models \cite{openai}. These embeddings capture semantic relationships between financial terms, enabling task-specific similarity mapping. Metadata, schema structures, and domain-specific keywords are indexed and stored in the vector database, optimized using scalable indexing algorithms such as Hierarchical Navigable Small Worlds (HNSW) \cite{hnsw}. This ensures that the system can rapidly match semantically related metadata fields with user intent at query time while supporting periodic updates from live financial glossary systems.

The Online Retrieval Service translates natural language queries into contextual vector representations, compares them against the indexed metadata vectors, and retrieves the most relevant matches in real-time. The service integrates a semantic reasoning layer that interprets financial terminology, such as "quarterly average balance" or "capital expenditure variance", ensuring that even ambiguous or non-technical queries yield precise results. It uses hybrid retrieval combining dense vector search for semantic matching with sparse keyword matching for accurate refinement using systems with vector search capabilities. Query outputs include ranked results, related metadata summaries, and when integrated with Large Language Models, an explanation of the retrieved information. Together, the Offline Indexing and Online Retrieval Services create a continuous feedback loop where retrieved insights inform future embeddings, leading to progressively smarter and context-aware financial search experiences.

  \begin{figure}[h]
  \centering
  \includegraphics[width=\linewidth]{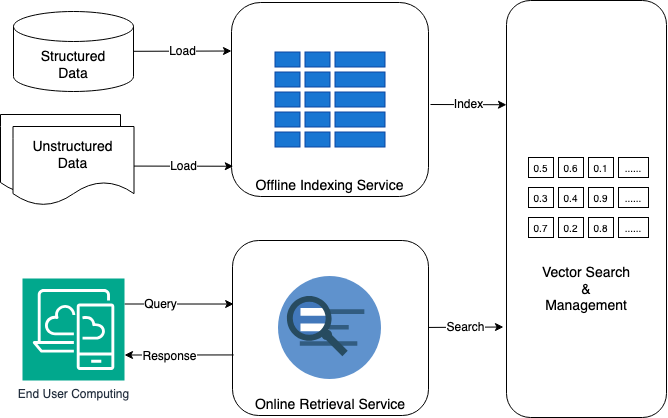}
  \caption{The NLQ System Overview}
  \end{figure}

\section{Offline Indexing Service}
Offline Indexing service involves curating data for natural language search; enabling fast and accurate query responses by performing intensive analysis ahead of time. This service monitors sources for updates, collects the data, and indexes the data to make it ready for retrieval. This process is generally run offline, which means it organizes and structures massive datasets outside the live or real-time environment, before it becomes available for user query. This indexing is performed in batch jobs as scheduled processes, typically using dedicated resources, and involves analyzing, parsing, and storing metadata, tokens, or vectors that enable fast and relevant search retrieval. 

Our proposed blueprint for Indexing Service is composed of 4 pipelines as described below. To meet scalability needs, each of the pipeline has been positioned to run on a Elastic Container Service (ECS) \cite{ecs} Cluster of Amazon Web Services.

\textbf{a. Frontier Pipeline:}
The first step should configure frontier pipeline consisting of a modular framework that, as described in Figure 2, spawns concurrent workers to efficiently automate queuing of objects from multiple sources. It starts by fetching system details by reading a configuration file or sourcing them from a specified web URL. This configuration should define the list of object identifiers, their source endpoints, authentication credentials and optional metadata. 

The pipeline should then parse the configuration to build an internal queue of fetch tasks which are pushed into an Extract Queue. Each task must include source details and mapping instructions that should define how the data should be structured once extracted. The pipeline’s design must emphasize scalability, allowing robust queuing of objects and flexible scheduling of the processes. Such setup allows us to seamlessly add any new knowledge source as required.
  \begin{figure}[h]
  \centering
  \includegraphics[width=\linewidth]{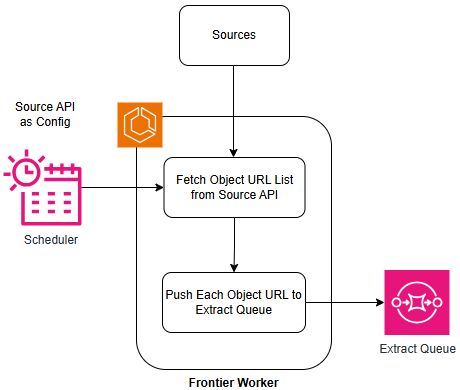}
  \caption{The Frontier Pipeline}
  \end{figure}

\textbf{b. Extract Pipeline:}
The next step is the Extract Pipeline that should spawn concurrent workers to efficiently pull each object from its designated data source, such as APIs, databases or cloud storage. It should maintain an Object Metastore to store the object hash and support the detection of object modification. Each worker should retrieve the object, calculate its hash and if this hash is different from the corresponding value in Object Metastore, it should then perform optional validation, push the extracted data into an Embedding Queue and update the object hash in the Metastore. This is described in detail in Figure 3.

This Embedding queue acts as a buffer for downstream processing of AI Enrichment pipelines. The configuration also allows fallback and retry policies, ensuring resilience when fetching data over unreliable networks.

  \begin{figure}[h]
  \centering
  \includegraphics[width=\linewidth]{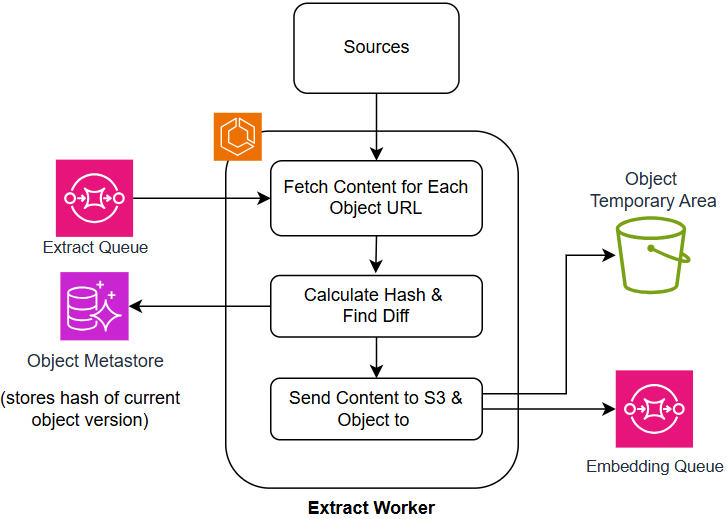}
  \caption{The Extract Pipeline}
  \end{figure}

\textbf{c. AI Enrichment Pipeline:}
The AI Enrichment Pipeline should consist of multiple services that call AI Models to enrich metadata of financial documents and datasets (like payment transactions, digital interactions, customer attributes, insurance documents, metadata catalogs, articles, news feeds) with additional attributes. This would improve the accuracy and relevance of downstream analytics and search operations. The pipeline should configure below two types of workers for Financial Metadata enrichment as outlined in Figure 4.

\textbf{c.1 Embedding Worker:}
Embedding workers extract the text from embedding queue and invoke a text embedding model to generate embeddings. These embeddings transform the financial metadata into dense numerical vectors that capture semantic relationships, enabling the system to compute similarity based on meaning rather than exact keywords. We used Amazon Bedrock's \cite{bedrock} amazon.titan-embed-text-v2 for our exprimentations. The generated vector data is stored in the designated object temporary area in S3 buckets and the metadata is passed to the Entity Tagging Queue for next stage of processing.

\textbf{c.2 Entity Tagging Worker:}
Entity Tagging Workers use natural language processing techniques called named entity recognition (NER) to tag entities/keywords to different objects like documents, metadata catalogs, articles and news feed. The entities tagged could be financial entities like payment system, home loan system or a brokerage tool or they could also be non-financial entities like a person (e.g. John Smith), an organization (e.g. Nvidia) or a place (e.g. Singapore). The processed object is written to Index Queue for further processing.

  \begin{figure}[h]
  \centering
  \includegraphics[width=\linewidth]{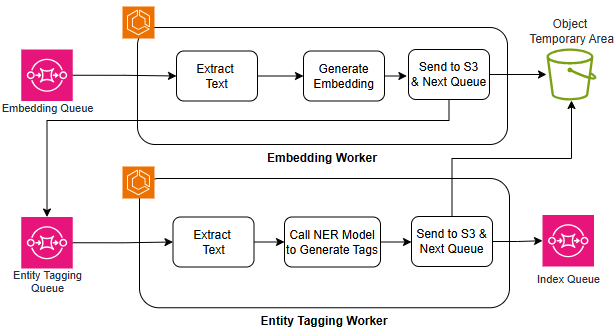}
  \caption{The AI Enrichment Pipeline}
  \end{figure}

We had access to a range of embedding models, as highlighted in Table 1, and collected the following observations during our experimentation phase before choosing the amazon.titan-embed-text-v2 model for our work. 

    \begin{table}[h]
    \begin{center}
    \begin{tabular}{ |>{\raggedright\arraybackslash}p{1.5cm}|>{\raggedright\arraybackslash}p{1.7cm}|>{\raggedright\arraybackslash}p{4cm}|  }
    \hline
    \textbf{Provider} & \textbf{Model} & \textbf{Observations} \\ 
    \hline
    Amazon & amazon. titan-embed-text-v2 & 8,192 token size; minimal rate limiting; less dimensional output than OpenAI, however satisfactory for financial documents \\
    \hline
    OpenAI & text-embedding-3-small & 8,192 token size; less dimensions compared to large version \\
    \hline
    OpenAI & text-embedding-3-large & 8,192 token size; better performance during execution compared to small version \\
    \hline
    Cohere & embed-english-v3.0 & token size too restrictive around 512 \\
    \hline
    \end{tabular}
    \caption{Embedding model observations}
    \end{center}
    \end{table}

\textbf{d. Index Pipeline:}
The Index Pipeline should be designed for efficient loading and search of vector data within a vector store, leveraging state-of-the-art indexing algorithms to optimize both scalability and search latency. We used Amazon's OpenSearch \cite{opensearch} as the vector store for our experiments. As the vectors are ingested, OpenSearch automatically constructs internal indexing structures based on configured algorithms for approximate nearest neighbor (ANN) search, such as Hierarchical Navigable Small Worlds (HNSW) and Inverted File System (IVF). For subsequent queries, our pipeline supports high-performance search over vector embeddings using k-Nearest Neighbor (k-NN) and scoring methods like cosine similarity or dot product. The process is described in Figure 5.

For NER data, it is beneficial to add keywords to the document for identified entities. OpenSearch will then be able to perform keyword-based search using the Okapi BM25 algorithm, which will rank documents based on the frequency and relevance of keywords present in each document relative to the query.

  \begin{figure}[h]
  \centering
  \includegraphics[width=\linewidth]{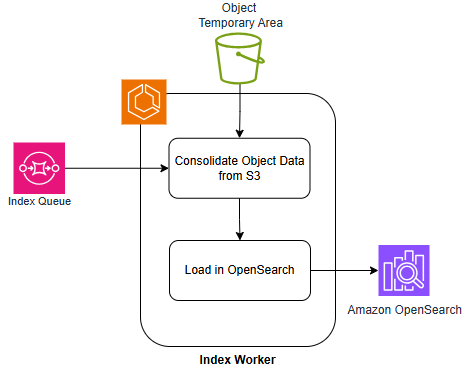}
  \caption{The Index Pipeline}
  \end{figure}

\section{Online Retrieval Service}

Even after best-in-class offline indexing, our systems will struggle to quickly and accurately find the right information if retrieval techniques are not equally optimized. The foundational keyword-based search systems focusing on retrieval of information and finding a query’s keywords and their existence within the indexed and/or tagged documents are no longer sufficient. Such processes fail to meet the complex demands for organizations requiring systems that can aid in regulatory compliance, risk assessment, and strategic decision-making across vast repositories of structured and unstructured data.

Traditional keyword searches are performed using entity tags and raw text, where a query’s words are searchable for matches within the document. For example, searching for “financial risk management” would return documents containing any of those words. The focus is on finding exact matches between query and document as can be seen in the Figure 6. In the past, many search engines including those embedded in websites, focus on finding keywords mentioned within a query and their existence within the indexed and/or tagged documents. Its next iteration was fuzzy search, which considered for other variations within the query string such as spellings and typos. This enhanced search approach is focused on increasing the likelihood of finding relevant results, despite different, incorrect or incomplete spelling. 
  \begin{figure}[h]
  \centering
  \includegraphics[width=\linewidth]{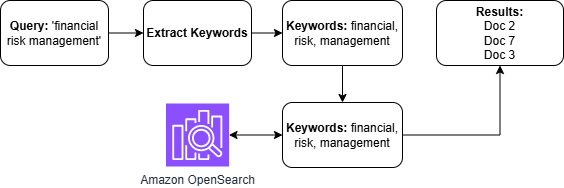}
  \caption{Keyword Search}
  \end{figure}

A more thorough, precise and equally fast process is required to meet such requirements. Apart from keyword-based search, many retrieval systems utilise semantic search by leveraging the embeddings of documents. The semantic search considers the underlying meaning behind a query as part of the search process. The system then performs a comparison of the vector embeddings generated by the query and the embedded knowledge base, with the aim to use different algorithms to facilitate vector-based retrieval. For the query “find three different credit assessment techniques” - the approach taken during the semantic search would involve understanding the intent behind the query and return the results that best match as visualised in Figure 7.
  \begin{figure}[h]
  \centering
  \includegraphics[width=\linewidth]{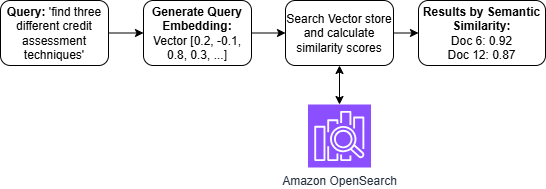}
  \caption{Semantic Search}
  \end{figure}

Given the requirement of regulatory compliance, risk assessment and strategic decision-making across vast repositories of structured and unstructured data, a further versatile searching methodology can prove more effective for financial services. Search methodologies with a singular angle fall short as both have a single retrieval approach, poorly handle infrequent terms and maintain limited ranking signals.

Hybrid search combines the perceived benefits of keyword-based searching and semantic searching, and seeks to address the limitations of each. Keyword (sparse) search guarantees that abbreviations, annotations, names, and code stay on the radar, while vector (dense) search ensures that search results are contextually relevant. The results are combined to provide a weighted score to ensure the top-k results have consistency across both measures. A popular example of where hybrid search has been implemented is Spotify's AI Playlist \cite{spotify}. 

  \begin{figure}[h]
  \centering
  \includegraphics[width=\linewidth]{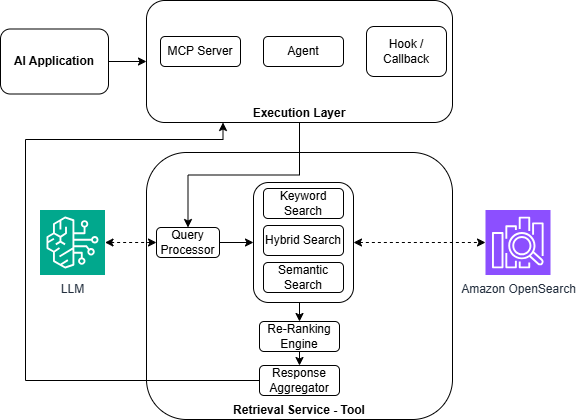}
  \caption{Retrieval Service}
  \end{figure}

Figure 8 is a proposed technical implementation of a comprehensive retrieval service that can help address limitations of singular methods. The structure of the service consists of a multi-faceted tooling that can handle different requirements, which may come from a range of financial service domains. Keyword, semantic and hybrid searches are considered alongside re-ranking to improve the validity of the results. For agentic systems, the tooling can be implemented under three methodologies: via Model Context Protocol (MCP), given as a direct tool for an agent to invoke, or as a callback/hook within an agentic workflow. Such retrieval can be defined in below 3 step process:

\begin{enumerate}
\item Query Reception \& Validation - AI Application receives and validates user query
\item Retrieval Tool Execution - System performs retrieval via execution layer (MCP, agent, or callback)
    \begin{enumerate}[label=(\alph*)]
    \item Query Processing - Validate input, generate embeddings, and determine search type
    \item Database Query - Execute search against knowledge store per configured strategy
    \item Results Processing - Validate and re-rank results for relevance
    \item Output Formatting - Aggregate and structure response payload
    \end{enumerate}
\item LLM Output Generation - Feed tool results into LLM to generate final response for user
\end{enumerate}

Our experimentation largely consisted of comparing the different search approaches under the same inputs as detailed in Table 2:
    \begin{table}[h!]
    \centering
    \begin{tabularx}{\columnwidth}{|>{\raggedright\arraybackslash}p{2cm}|>{\raggedright\arraybackslash}X|}
    \hline
    \textbf{Query} & \textbf{Observations} \\ \hline
    `Financial services risk management' & KW: < 10s\newline SEM: >10s\newline HYB: >30s\newline A straightforward request with only keywords. Keyword search was fastest and aligned with output expectations. \\ 
    \hline
    `What are some of the known facts about customer acquisition tables?' & KW: < 10s\newline SEM: >20s\newline HYB: >30s\newline A more generic query, likely to be asked by an analyst. The observation was that despite the different timings, the outcomes were comparable and close to expectation for known tables. \\
    \hline
    `Uh, so like, how do financial services, you know, manage risks and all that?' & KW: < 10s\newline SEM: >20s\newline HYB: >30s\newline The addition of extra filler words represents the common issue where keyword search begins to struggle. In this scenario, despite taking the longest time, hybrid search with re-ranking enabled produced the most relevant results. \\ 
    \hline
    \end{tabularx}
    \caption{Retrieval service observations}
    \end{table}

This comprehensive online retrieval service architecture, combining hybrid search methodologies with intelligent re-ranking and flexible deployment options, provides financial service institutions with the robust yet adaptable information discovery capabilities essential for navigating complex regulatory landscapes and extracting actionable insights from their diverse data ecosystems.

\section{Applications in Real World}

The practical deployment of NLQ system design in financial services environments can unlock transformational applications across many critical domains:

\textbf{a. Financial Analytics through Text-to-SQL Conversion:} The ever-growing operating and reporting requirements that large financial services face lead to traditional data storage systems hitting bottlenecks. With many tasks like critical business processes, risk assessment and regulatory reporting now considered a non-negotiable, text-to-SQL capabilities can help democratise the data for domains, enabling processing of natural language requests. Via the index service, data can be properly indexed and catalogued, ensuring that the retrieval service can translate the natural language into optimised queries that become the bridge between business questions and database execution.

\textbf{b. Retrieval-Augmented Generation for Knowledge Synthesis:} Financial services need to manage a vast repositories of documentation, policies and market intelligence for the purpose of decision-making. RAG applications can help navigate the knowledge at scale which includes aiding compliance teams interpret current regulations, supporting investment analysts synthesize insights from multiple research sources or helping relationship managers assess their client history in an efficient manner. The index service can be setup to continuously ingest documents along with the required enrichment and the retrieval service can be placed into the hands of the employees to perform searches to retrieve the most relevant context.

\textbf{c. Recommendation Systems for Content Discovery:} Being in a customer centric industry, the financial service institutions always seek to better understand and service their customers based on their past interactions. The index service can assist in building comprehensive user profiles via automated pipelines, which can then be utilised by the retrieval service to identify semantically similar documents to predict service needs based on current context and help predict their future behaviour. 

\textbf{d. ML Feature Store for Faster Model Development:} NLQ on top of ML feature stores can enable quantitative analysts to rapidly identify proven risk indicators, equip model validators to trace feature origins for regulatory documentation, and allow data scientists to discover existing features that prevent redundant development. This capability can accelerate time-to-market for new models, ensures regulatory compliance through comprehensive audit trails, and reduces operational risk through standardised feature definitions.

\textbf{e. Metadata Augmentation for Enhanced Data Governance:} The exponential growth in data generation and collection across financial institutions has intensified regulatory scrutiny and compliance requirements, with governments and financial regulators demanding greater transparency. Companies are facing the burden of manual actions and validations in ensuring they have comprehensive data governance across all their functions. Automated metadata enhancement can transform this compliance into an intelligent process that can cover classifying financial data, documenting business processes and maintain appropriate controls and audit trails. The index service can continuously and consistently enrich the data assets with automated sensitivity tagging and compliance classification, whereas the retrieval service can assist in discovering data relationships and lineage information in a self-service mode for the generation of compliance reports.
 
\section{Future Direction}

As the data continues to expand rapidly across all areas of financial services, there is a growing need for NLQ systems that are both powerful and adaptable to handle complex information retrieval efficiently. These systems must aid in the discovery of insights from vast, heterogeneous information repositories that span regulatory documents, market research, internal policies, and transactional records.

With regards to offline indexing service, the given blueprint can be expanded to explore more advanced embedding models like NV-Embed-v2 by Nvidia or  gte-Qwen2-7B-instruct by Alibaba to support better contextual search in future. By capturing deeper semantic relationships between words and concepts, these advanced embedding models will lead the design towards more accurate and meaningful retrieval results. There is also an opportunity to explore graph databases to enhance search system by directly storing and traversing relationships between data points, enabling faster and contextual retrieval of interconnected information. Graphs will enrich results by mapping entities, concepts and metadata into knowledge graphs, allowing the system to understand semantic connections and deliver more relevant, explainable, and human-like responses. 

By moving away from a world of traditional static information retrieval and adopting modern retrieval services, especially within the financial services landscape, we can look at below enhancements to the blueprint to meet the increasingly sophisticated demands of financial services organizations:

\textbf{a. Multimodal Integration:} This will ensure a cohesive understanding of the relationships between different artifacts of information to provide a more relevant response to queries, whether those are PDFs with charts, spreadsheets, documentation platforms, diagrams, or recordings.

\textbf{b. Real-Time Adaptive Learning:} Modern retrieval services should account for real-time adaptive learning focused on understanding the query at a deeper level, tuning the results through new forms of re-ranking and personalising the results based on personas of the users. 

\textbf{c. Domain-Specific Compliance \& Security:} Regardless of the size and complexity of the systems, the domain-specific compliance \& security needs to remain at the forefront. Advanced retrieval systems should seek to ensure access control on results, generate audit trails, explain rankings, and privacy treatment such as redaction of sensitive information.

\textbf{d. Agentic RAG:} We can look to introduce Agentic RAG capabilities to our architecture that can provide dynamic orchestration of search, retrieval, and reasoning steps to deliver context-driven and intelligent responses.

\section{Conclusion}

Building a modern NLQ system tailored for financial documents and datasets represents a critical advancement in addressing the complexities and demands of contemporary financial data analysis. By integrating state-of-the-art techniques in natural language processing, semantic understanding, and vector-based data linking, such systems enable more precise and context-aware access to financial information. This not only supports decision-makers in navigating a rapidly evolving financial landscape but also enhances the capabilities of regulatory compliance, market intelligence, and risk mitigation.

Through our experiments and reference implementation of this system design, we feel well-placed to make two key conclusions. Firstly, the structure of the raw data plays a significant role in embedding generation such that the extract and enrichment worker processes stay compatible with each other. Secondly, for the retrieval service, the weighting of hybrid search can dictate the relevance of results, and should be tuned based on the data source and domain of the user.

In the future, we would like to expand the NLQ system blueprint to support multi-modal inputs. This would help onboard a range of different data repositories in financial institutions and enable adaptive learning to increase the system's intelligence and flexibility. The systems and methods discussed in this paper set a foundation for future research that will refine search strategies, improve scalability, and consolidate the integration of diverse data sources. Ultimately, modern financial NLQ hold the promise of transforming how users discover, interpret, and leverage financial knowledge in an increasingly data-driven world.

\section{Generative AI Declaration}
We have used LLMs to help in editing of this paper which includes automated spell checks, word corrections and grammatical improvements.

\end{document}